\documentclass[namedreferences]{SolarPhysics}
\usepackage[optionalrh]{spr-sola-addons} 
\usepackage{graphicx}        
\usepackage{url}             

\newcommand{\etal}{{\it et al.}}
\newcommand{\eg}{{\it e.g.}}
\newcommand{\ie}{{\it i.e.}}

\newcommand{\adv}{    {\it Adv. Space Res.}}

\newcommand{\aap}{    {\it Astron. Astrophys.}}

\newcommand{\apj}{    {\it Astrophys. J.}}
\newcommand{\apjl}{   {\it Astrophys. J.}}
\newcommand{\apjs}{   {\it Astrophys. J. Suppl. Ser.}}

\newcommand{\jastp}{  {\it J. Atmos. Sol. Terr. Phys.}}
\newcommand{\jgr}{    {\it J. Geophys. Res.}}

\newcommand{\nat}{    {\it Nature}}

\newcommand{\pasj}{   {\it Pub. Astron. Soc. Japan}}

\newcommand{\solphys}{{\it Solar Phys.}}

\begin{document}
\begin{article}
\begin{opening}

\title{Reconnection of a kinking flux rope triggering the ejection of a
       microwave and hard X-ray source}
\subtitle{II. Numerical Modeling}

\author{B. \surname{Kliem}$^{1,2,3}$,
        M. G. \surname{Linton}$^{3}$,
        T. \surname{T{\"o}r{\"o}k}$^{4}$,
        M. \surname{Karlick\'y}$^{5}$}
        \runningauthor{Kliem et al.}
        \runningtitle{Reconnection in kinking flux rope ejection II}

\institute{$^{1}$ Universit\"{a}t Potsdam, Institut f\"{u}r Physik und
                  Astronomie, 14476 Potsdam, Germany
                  \email{bkliem@uni-potsdam.de}\\
           $^{2}$ University College London, Mullard Space Science Laboratory,
                  Holmbury St.~Mary, Dorking, Surrey, RH5 6NT, UK\\
           $^{3}$ Naval Research Laboratory, Space Science Division,
                  Washington, DC 20375, USA\\
           $^{4}$ LESIA, Observatoire de Paris, CNRS, UPMC, Universit\'e Paris
                  Diderot, 5 place Jules Janssen, 92190 Meudon, France\\
           $^{5}$ Astronomical Institute, Academy of Sciences of
                  the Czech Republic, 251 65 Ond\v{r}ejov, Czech Republic}

\date{Received 24 March 2010; accepted 7 July 2010}

\begin{abstract}  Numerical simulations of the helical ($m\!=\!1$) kink
instability of an arched, line-tied flux rope demonstrate that the helical
deformation enforces reconnection between the legs of the rope if modes with
two helical turns are dominant as a result of high initial twist in the range
$\Phi\gtrsim6\pi$. Such reconnection is complex, involving also the ambient
field. In addition to breaking up the original rope, it can form a new,
low-lying, less twisted flux rope. The new flux rope is pushed downward by the
reconnection outflow, which typically forces it to break as well by
reconnecting with the ambient field. The top part of the original rope, largely
rooted in the sources of the ambient flux after the break-up, can fully erupt
or be halted at low heights, producing a ``failed eruption.'' The helical
current sheet associated with the instability is squeezed between the
approaching legs, temporarily forming a double current sheet. The leg-leg
reconnection proceeds at a high rate, producing sufficiently strong electric
fields that it would be able to accelerate particles. It may also form
plasmoids, or plasmoid-like structures, which trap energetic particles and
propagate out of the reconnection region up to the top of the erupting flux
rope along the helical current sheet. The kinking of a highly twisted flux rope
involving leg-leg reconnection can explain key features of an eruptive but
partially occulted solar flare on 18 April 2001, which ejected a relatively
compact hard X-ray and microwave source and was associated with a fast coronal
mass ejection.

\end{abstract}


\end{opening}

\section{Introduction}
\label{s:introduction}

When a line-tied, arched magnetic flux rope becomes unstable with
respect to the helical kink mode, \ie, to displacements with azimuthal
wavenumber $m=1$, several effects occur which are relevant to the
understanding of solar eruptions \cite{Sakurai1976, Fan&Gibson2003,
Torok&al2004}. First, the apex of the rope will ascend if the
initial perturbation points upward, which is usually the case for solar
eruptions due to a slow rise of the structure prior to the main
acceleration. Second, the ensuing fast rise is initially exponential or
a power law close to an exponential, as observed to be typical for fast
coronal mass ejections (CMEs) \cite{Vrsnak2001, Schrijver&al2008a}.
Third, the rope will writhe by the conversion of twist helicity into
writhe helicity. This leads to a rotation of the apex about the
direction of ascent and forms a characteristic inverse gamma shape or O
shape (depending on the perspective of the observer). Such shapes are
often observed and have been taken as evidence for the occurrence of the
instability (\eg, \opencite{Rust2003}; \opencite{Ji&al2003};
\opencite{Romano&al2003}; \opencite{Rust&LaBonte2005};
\opencite{Torok&Kliem2005}; \opencite{Fan2005}; \opencite{Zhou&al2006};
\opencite{Gilbert&al2007}; \opencite{Cho&al2009}).
Fourth, two current sheets will be formed and exponentially steepened,
so that the onset of magnetic reconnection is likely: a vertical
(flare) current sheet under the rope resulting from the rope's rise;
and a helical current sheet, wrapped around the rope and passing over
the rope's apex, created as a consequence of the helical displacement.
Reconnection in these current sheets influences the eruption
strongly. It leads to a variety of different
dynamical behaviors, including ejective and confined eruptions
\cite{Torok&Kliem2005} as well as the break-up and reformation of the
rope, which we will address in the present paper. Fifth, sigmoids are
likely to form, since the field lines that thread either of the current
sheets are sigmoidal in projection \cite{Titov&Demoulin1999,
Fan&Gibson2003, Kliem&al2004, Gibson&al2004}. The orientation of the
sigmoidal field lines underneath the flux rope, where sigmoids are known
to form, correlates with the chirality of the field in agreement with
the observations \cite{Rust&Kumar1996, Pevtsov&al1997, Green&al2007}.

The literature reports a number of events that suggest the occurrence of the
helical kink instability through observational evidence for both the existence
of considerable twist (exceeding one field line turn about the axis of the
structure) and the development of writhe in the course of the eruption.
Examples are the events of 19 July 2000 \cite{Romano&al2003}, 18 August 1980
\cite{Vrsnak&al1993}, and 12 June 2003 \cite{Liu&Alexander2009}; see also the
systematic study by \inlinecite{Vrsnak&al1991}.

In spite of such considerable principal agreement and support, it is
possible that the instability actually occurs less frequently than
indicated by the above observations, because there exists a second
physical process that can cause the writhing of a rising flux rope into
helical shape. As has been pointed out by
\inlinecite{Isenberg&Forbes2007}, the shear field component of the
ambient field (pointing along the photospheric polarity inversion line),
crossed with the vertical current component in the legs of a flux rope
which is displaced upwards out of equilibrium, gives a sideways pointing
Lorentz force on the flux rope legs. This force yields a writhing in the
same direction as the kink-mode-driven writhing for \emph{any} process
that lifts the rope out of equilibrium. Basically, the same
observational consequences as listed above can be expected, including
the reduction of any possibly existing initial twist into the developing
writhe, due to helicity conservation.

Therefore, to evaluate the relevance of the helical kink instability for
solar eruptions, it is important to distinguish between the two possible
causes of the writhing into helical shape. The direct
approach---computing the coronal field by extrapolation of vector
magnetograms and comparing the obtained twist to the threshold of the
helical kink mode---is still hardly practicable because present vector
magnetograms and nonlinear force-free extrapolation codes do not appear to
have reached the degree of consistency required for a reliable
computation (\opencite{Metcalf&al2008}, \opencite{Schrijver&al2008b}),
and because a comprehensive parametric study of the instability
threshold has not yet been completed (although some knowledge is already
available, see \opencite{Torok&al2004} and the references therein).
Hence, it is of interest to identify features that permit an
observational discrimination.

The writhing by the shear field component has the property
that the sideways pointing Lorentz force acting on the flux rope legs
has the same sign all along one leg, and the opposite sign along the
other leg; it changes sign only at the
apex. This implies that the displaced flux rope cannot develop more than
one helical turn. In terms of a normalized axial wavenumber,
$k'=kl/2\pi=l/\lambda$, where $l$ is the length of the flux rope and
$\lambda$ is the wavelength of the perturbation, this can be expressed
as $k'\le1$.

In contrast, the helical kink instability can have any
wavenumber in principle. Its dominant wavenumber is set by the twist
$\Phi=lB_\phi/rB_z=2\pi N$, where $N$ is the number of field line turns
in the flux rope, in general averaged over its cross
section. Stability analyses indicate that maximum growth occurs in the
range $\lambda\sim(1\mbox{--}2)l_\mathrm{p}$, where $l_\mathrm{p}=l/N$
is the pitch length, the axial length required for one field line turn
\cite{Linton&al1996, Linton&al1998}. For a force-free flux rope with
uniform radial twist profile, for example, maximum growth occurs for
$\lambda=2l_\mathrm{p}$. In this case, the $k'\sim1$ and $k'\sim2$ modes
are dominant for $\Phi\sim4\pi$ and $\Phi\sim8\pi$, respectively, so
that the transition between them
occurs in this range of twist values. Such high values render the
occurrence of $k'>1$ modes significantly less likely than the occurrence
of single helical modes. For twist profiles that lead to maximum growth
closer to $\lambda=l_\mathrm{p}$, the transition will occur between
twists of $2\pi$ and $4\pi$, \ie, closer to the kink instability
threshold, so it could be more common. However, it is not clear whether
such nonuniform profiles are often formed in the solar atmosphere. The
observations of erupting filaments rather tend to indicate $k'\lesssim1$
helical modes in the majority of events.

If modes $k'>1$ are observed, the helical kink instability is implied.
Projected onto the plane of the sky, this manifests itself as the occurrence of
more than one inflection point of the kinked flux tube. For events near Sun
center, an S shape with an overturn at one or both ends results (see the
kinking flux rope in Figure~\ref{f:rope_fl1} below at $t=11$ for an
illustration). In fortunate cases the effect is apparent even in limb events,
for example in the failed filament eruption of 27 May 2002, which showed an
inflection point in the middle of the front leg (see the middle panels of
Fig.~1 in \opencite{Torok&Kliem2005}). If $k'$ is clearly larger than unity,
then the multiple helical shape is relatively clearly indicated for any
perspective; an example is the 19 July 2000 filament eruption analyzed in
\inlinecite{Romano&al2003}. Such direct observations of the helical mode with
$k'>1$ are not very numerous, however.

Another possible consequence of $k'>1$ mode dominance is an approach of the
rising flux rope's legs so that they can come into contact and begin to
reconnect. Such an evolution is not triggered by a $k'\le1$ helical
displacement, which rather causes the legs of the flux rope to move apart. It
is expected to have distinct observational consequences. The reconnection can
cause the break-up of the original flux rope as suggested by
\inlinecite{Cho&al2009}. Furthermore, the simulations in Section~\ref{s:hKI}
demonstrate that such break-up can lead to the reformation of a (less twisted)
flux rope from the two bottom parts of the original rope, which facilitates the
quick reformation of a filament after an eruption. Finally, the energetic
particles produced by the reconnection can have unique radiation signatures if
the helical current sheet is involved. This was found in the companion paper to
the present study (\opencite{Karlicky&Kliem2010}; hereafter Paper~I), which
analyzes the observations of an eruptive flare on 18 April 2001 in NOAA Active
Region (AR) 9415. This partly occulted eruption formed a rapidly ascending
microwave source of strongly helical (inverse gamma) shape (shown also in
Figure~\ref{f:comparison1} below) and presents evidence for the acceleration of
nonthermal particles at or near the crossing point of the flux loop legs (as
seen in the projection onto the plane of the sky). Moreover, the particles were
efficiently trapped in superimposed moving compact sources, which are
suggestive of plasmoids formed by reconnection between the legs and moving in
the helical current sheet.

In the present paper, we demonstrate the reconnection between the legs of a
kinking flux rope using magnetohydrodynamic (MHD) simulations. We estimate the
minimum twist enabling this process and study its dynamics. The simulation
results are then discussed in relation to the observations of the eruptive
flare on 18 April 2001 which were found in Paper~I to be suggestive of such
reconnection.

\section{Helical Kink Instability of a Highly Twisted Flux Rope}
\label{s:hKI}

\begin{figure}[t]                                                
\centering
\includegraphics[width=.8\textwidth]{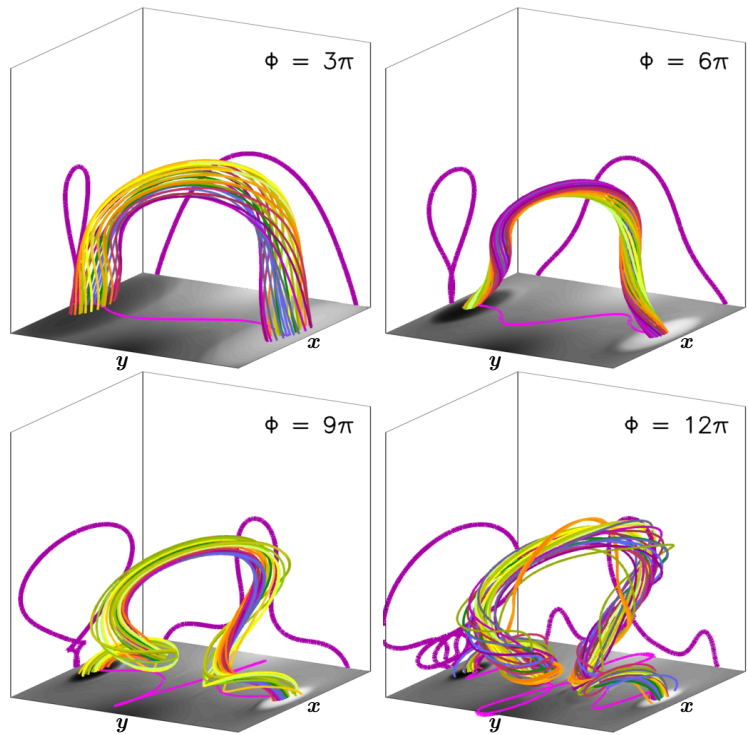}
\caption{
Flux rope shape for a range of initial twist values and uniform apex
height of twice the initial apex height (reached at $t=34$, 19, 13, and
$12\tau_A$ in the simulations). The volume $1\times4\times4$ is shown; the
stretching in the $x$ direction helps to visualize the helical shapes.
All field lines are started from circles of radius $0.2a$ centered at
the footpoints of the flux rope. The orthogonal projections show the
magnetic axis of the rope.}
\label{f:shape(pi)}
\end{figure}

The zero-beta, ideal MHD simulations presented in this paper are all
largely similar to those in \inlinecite{Torok&Kliem2005}. They use the
force-free equilibrium of a toroidal flux rope by
\inlinecite{Titov&Demoulin1999} as initial condition and employ similar
numerical parameters (a Cartesian box of size $10\times10\times20$,
discretized by a stretched grid of resolution $\Delta=0.03$ in the
central part, and very similar small viscosity and weak spatial
smoothing, as required to maintain numerical stability). Magnetic
reconnection occurs in these simulations as a consequence of numerical
diffusion where current layer(s) steepen in response to the development
of ideal MHD instability.

Most geometrical parameters of the rope, as well as the normalization,
are also set as in these previous simulations: major torus radius
$R=1.83$, depth of torus center $d=0.83$, distance of the sources of the
external poloidal field (the field components perpendicular to the
flux rope axis) $L=0.83$. The line current in the model, which produces
the external toroidal field, is reduced by a factor 20 in comparison to
the simulation of a failed eruption in that previous paper, to enable
eruptive behavior \cite{Roussev&al2003, Torok&Kliem2005}. The
toroidal field component points along the flux rope axis and is
essentially also the shear field component, since the polarity inversion
line of the field in the bottom plane runs approximately parallel to the
flux rope under its apex. The external field is thus mainly poloidal,
with the ratio between toroidal and poloidal components at the flux rope
apex being $B_\mathrm{et}/B_\mathrm{ep}=0.075$. Consequently, any
development of writhe is nearly exclusively due to the helical kink
mode. Gravity is neglected, and the initial density is derived from the
initial magnetic field by $\rho_0=|\mathbf{B}_0(\mathbf{x})|^{3/2}$, to
model a corona with an Alfv\'en velocity that decreases slowly with
distance from the flux concentrations. Initially, the flux rope lies in
the plane $\{x=0\}$, and its field strength is normalized to unity at
the apex, $B_0=|\mathbf{B}(0,0,1,0)|=1$. This gives an initial
Alfv\'en velocity $V_{A0}=B_0\rho_0^{-1/2}=1$ at the apex, which serves
as the velocity unit and defines the time unit through
$\tau_A=1/V_{A0}$.

The average twist of the flux rope is varied between the simulations
in the range $\Phi=(3\mbox{--}16)\pi$ (in steps of $1\pi$ between 3 and
$10\pi$) by varying the minor torus radius between $a=0.60$ and
$a=0.11$. All these equilibria are unstable with respect to the helical
kink mode. See \inlinecite{Torok&al2004} for the calculation of the
average twist, for the overall evolution of the instability in such
equilibria, in particular for the formation of the flare current sheet
and the helical current sheet, and for a parametric study of the growth
rate.

For equilibria of progressively higher twist, the axis of the
kinked flux rope shows progressively higher values of the number of
helical turns, with the transition to a dominance of the double
helical mode ($k'\sim2$) occurring for $\Phi\approx6\pi$. As a
consequence, the helical deformation of the rope legs leads to a
progressively smaller separation between the legs under the rising rope
apex (Figure~\ref{f:shape(pi)}). This facilitates the onset of
reconnection between the legs. For much higher twists
($\Phi\gtrsim12\pi$), higher axial modes ($k'\gtrsim3$) dominate
initially, but the upward expansion of the flux rope leads to the
dominance of the double helical perturbation in the nonlinear evolution
of the instability, an effect we will address in a future study.

The twist value at the transition between the regimes of single and
double helical mode dominance depends not only upon the radial twist
profile of the rope as discussed in the Introduction, but also on the
height profile of the overlying potential field. This dependence will be
studied in more detail in a future paper. Our finding of
$\Phi\gtrsim6\pi$ for double helical mode dominance may be compared with
the simulation of the helical kink instability in a helmet streamer-like
configuration above a highly twisted flux rope in
\inlinecite{Birn&al2006}. The erupting top part of their flux rope had
twists in the range $\approx\!(9\mbox{--}12)\pi$ and it clearly showed
the double helical shape of a $k'\approx2$ perturbation with an axis
shape similar to the $\Phi=9\pi$ snapshot in Figure~\ref{f:shape(pi)}.
Unfortunately, no statement was given whether the legs did begin to
reconnect in the later phases of the evolution.

\begin{figure}[ht!]                                                
\centering
\includegraphics[width=.5\textwidth]{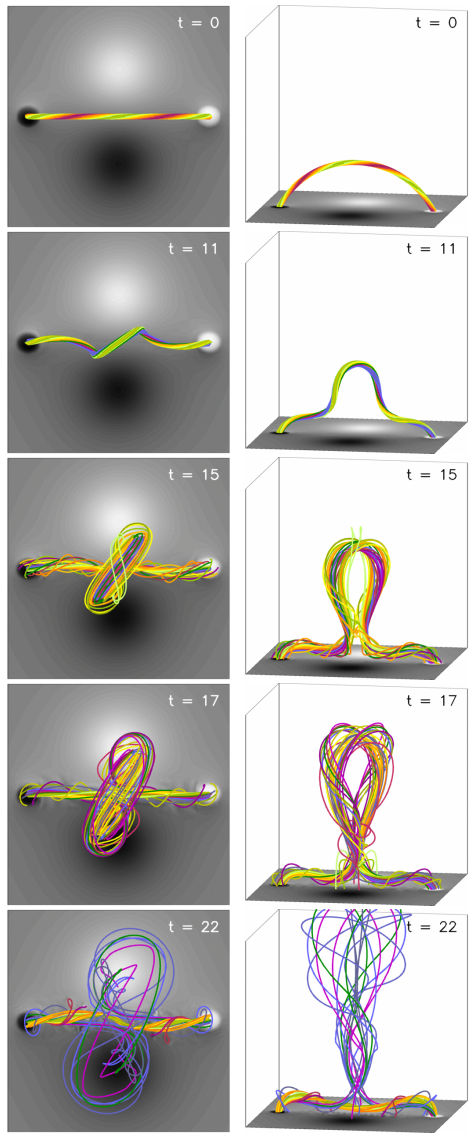}  
\caption{
Field lines in the core of a kink-unstable flux rope with initial
average twist $\Phi=9\pi$ (started at circles of radius $r=0.2a$
centered at each footpoint of the magnetic axis of the rope) in top and
perspective views. This shows the evolution of the double helical
($k'\sim2$) perturbation, which leads to the approach and subsequent
reconnection of the flux rope legs. A new low-lying, less twisted flux
rope is formed (bottom panels). The inner part (of size
$4\times4\times4$) of the simulation box is displayed, including the
magnetogram, $B_z(x,y,0,t)$.}
\label{f:rope_fl1}
\end{figure}

\begin{figure}[t]                                                
\centering
\includegraphics[width=\textwidth]{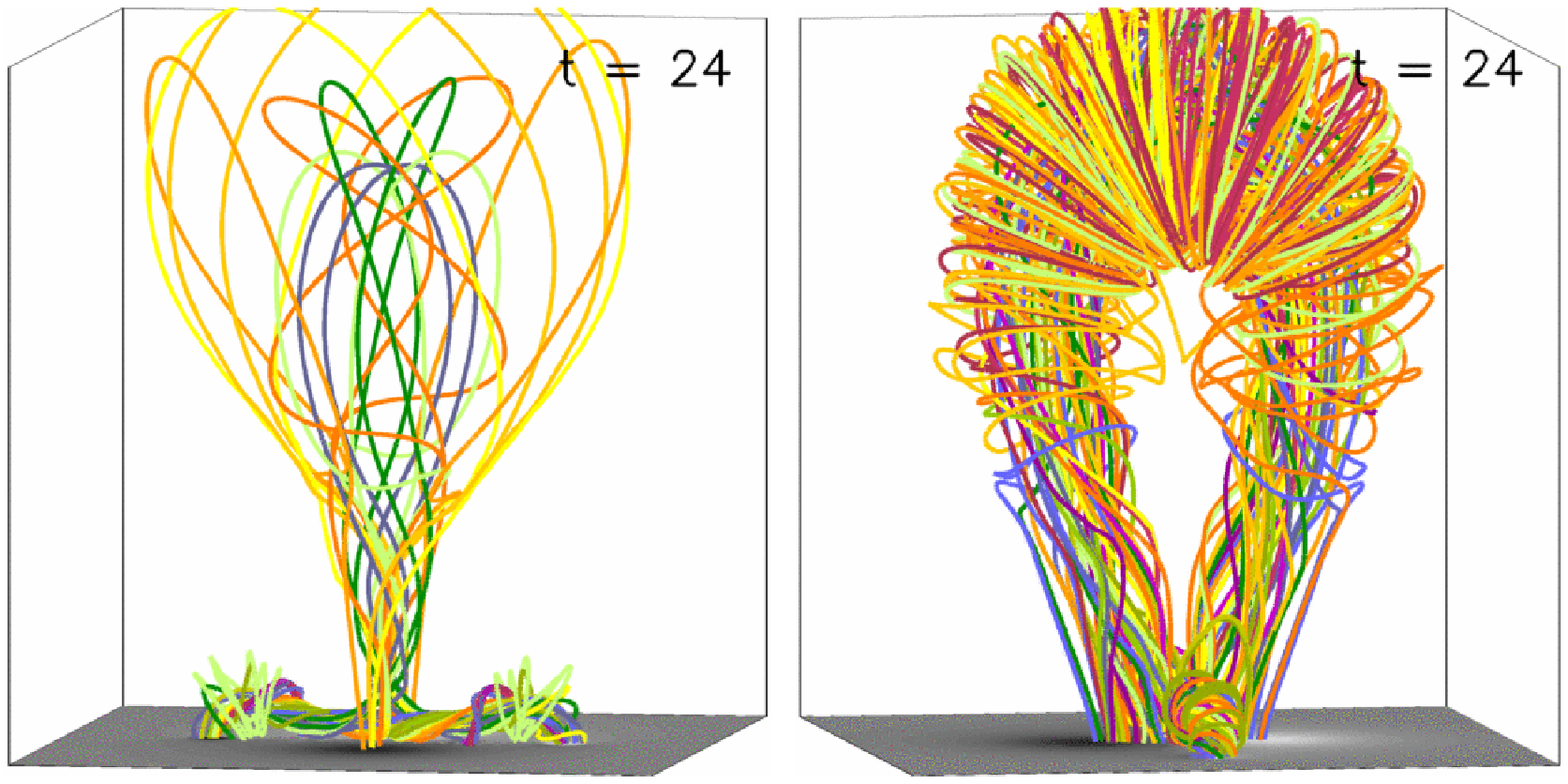}  
\caption{
{\it Left:} Field lines started at the same positions as in
Figure~\ref{f:rope_fl1} at a later time, after the original flux rope
legs have reconnected. A new flux rope is formed below the reconnection
region, which remains at low heights ($z\lesssim1$), while the top part
of the original rope continues the upward expansion.
{\it Right:} Field lines started in the top part of the rising flux at
the same time, showing the flux rope structure and the new magnetic
connections to the sources of the originally overlying flux. This is
supplemented by some field lines started at the flux rope footpoints.
The inner subvolume of size $6^3$ is shown. The views differ by a
rotation of $70^\circ$ about the vertical ($z$) axis.}
\label{f:rope_fl2}
\end{figure}

\subsection{Reconnection in the $m=1$, $k'\sim2$ Helical Kink}
\label{ss:R_hKI}

In the following we discuss one run of the series in more detail, taken to be
the run with $\Phi=9\pi$. This run shows a clear dominance of the double
helical mode, which we intend to relate to the observations of the 18 April
2001 eruption. While this event motivated our study, we do not attempt to model
the observed source shape, apex trajectory, and height of the crossing point
very closely in this section. The purpose of the simulation is
primarily to demonstrate that the legs of the rope reconnect if the twist is
sufficiently high, with the apex continuing to rise at high speed and the
region of leg-leg reconnection not participating in the fast rise.
Quantitative comparisons between simulation and observation are given
in Section~\ref{ss:hKI+TI} and in Paper~I.

Figure~\ref{f:rope_fl1} presents snapshots of field lines in the core of
the flux rope which show the initial condition, the evolution of the
double helical $k'\sim2$ kink perturbation, and the approach and
interaction of the rope legs. The reconnection of the outer flux layers
of the rope begins between the second and third snapshot pair and the
corresponding perturbation has propagated into the core of the rope by
the time of the third snapshot pair, distorting the field lines,
although the core still has the original field line connections. The
final snapshot pair shows the system when most of the flux in the
original rope has reconnected, resulting in the break-up of the original
rope and reformation of an only weakly kinked, still twisted flux rope
at low heights.

Both the rise of the rope's upper part and the reconnection at low
heights continue subsequently. Figure~\ref{f:rope_fl2} illustrates that
the rising flux still has the character of a flux rope, although its
average twist is reduced and although it becomes increasingly rooted in
the ambient flux, which is obviously involved in the reconnection that
breaks up the original rope. The downward reconnection outflow pushes
the newly formed flux rope toward the bottom boundary, enforcing further
reconnection with the ambient flux. Eventually, the newly formed rope
breaks up as well into two flux bundles that connect each footpoint of
the original rope to the corresponding external flux concentration of
opposite polarity (not shown); this is analogous to the corresponding
potential field.

The rise of the apex is initially exponential, reaching $0.3V_{A0}$ by
$t\approx12\tau_A$, followed by an approximately linear rise (with slowly
decreasing velocity), qualitatively similar to the two rise profiles shown
below in Figure~\ref{f:rise} (Section~\ref{ss:hKI+TI}). This is the typical
behavior of fast CMEs and consistent with the 18 April 2001 eruption, which
showed a fast linear rise after the occulted short main acceleration phase. We
note that the flux rope enters the height range of the torus instability in the
course of its main acceleration. This instability, which is a form of the
lateral kink \cite{Kliem&Torok2006}, eventually drives the flux rope to a full
eruption.

\begin{figure}[t]                                                
\centering
\includegraphics[width=.69\textwidth]{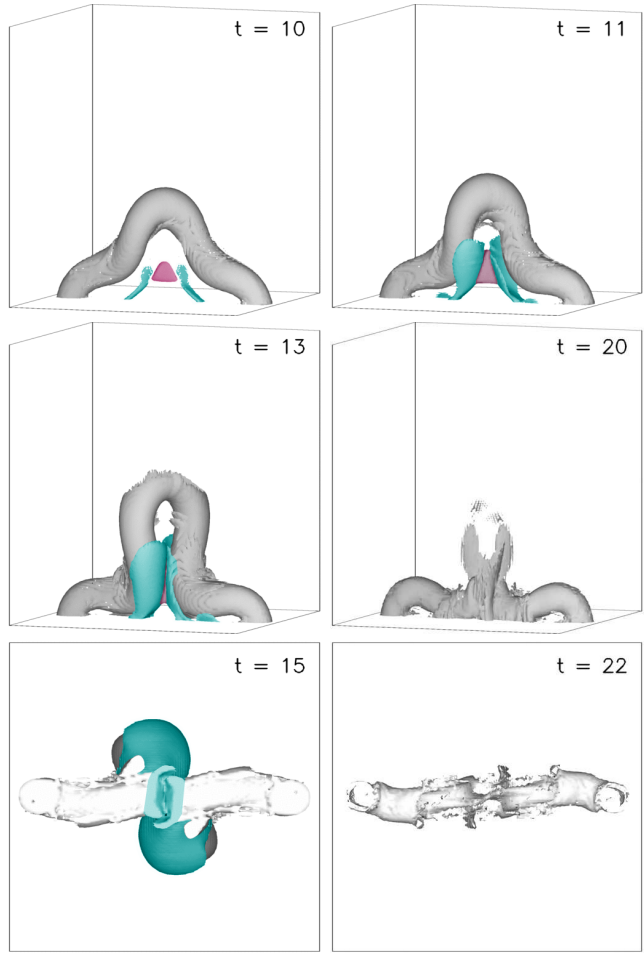}  
\caption{
Isosurfaces of current density, at 10\% of the peak current density in
each snapshot, showing the kinking, reconnection, and reformation of the
current channel (gray) in the core of the flux rope which is displayed in
Figures~\ref{f:rope_fl1} and \ref{f:rope_fl2}. In addition, the formation
of the vertical (flare) current sheet (red) and of the lower parts of the
helical current sheet (cyan) are apparent in panels~1 and 2. Panel~3 shows the
temporary formation of a double current sheet from two sections of the
helical current sheet between the approaching flux rope legs just before
the legs reconnect. Upward views through the bottom of the box in
panels~5 and 6 show the bottom edge of the squeezed current sheet (cyan)
between the light-gray horizontal sections of the flux rope legs, the
dark-cyan underside of the helical current sheet that wraps around the
upper part of the flux rope legs ($t=15$), and the current flow through
the reconnected horizontal sections of the flux rope legs ($t=22$). The
inner subvolume $4^3$ is displayed.}
\label{f:rope_iso_j}
\end{figure}

The isosurfaces of the current density plotted in
Figure~\ref{f:rope_iso_j} show the development of current sheets, in
addition to the kinking, interaction, and reformation of an
approximately straight current channel in the core of the flux rope. The
first panel displays the onset of current density steepening in the
vertical (flare) current sheet below the rising flux rope apex
(red triangular structure) and in the helical current sheet (two oblique
cyan layers near the knees of the bent channel). While the vertical current
sheet exists only below the rising current channel, the helical current
sheet actually wraps around the whole channel, passing over the apex.
Only its lower sections are visible in these isosurface plots, since the
current density in the helical sheet decreases rapidly with height as a
consequence of the rapidly decreasing flux density of the ambient field.
Therefore, the helical displacement of the upper parts of the flux rope
cannot pile up much flux in front of the interface to the ambient field,
where the helical current sheet forms. (See, \eg, Fig.~7 in
\opencite{Arber&al1999}, Fig.~4 in \opencite{Gerrard&al2001}, and Fig.~3
in \opencite{Torok&al2004} for isosurface plots that show the complete
helical current sheet formed around kinking flux ropes which are
embedded in less inhomogeneous field, and Fig.~10 in Paper~I for field
line plots of the helical current sheet. See also
\opencite{Lionello&al1998} for the formation of the helical current
sheet in a representative range of kink-unstable flux rope equilibria.)

The second and third panels in Figure~\ref{f:rope_iso_j} show how the
approaching legs of the kinking rope squeeze the two layers of the helical
current sheet and the vertical current sheet between them into a vertical
double current sheet. The fourth panel demonstrates that the double current
sheet and the legs of the kinking current channel reconnect, reforming a
low-lying current channel (flux rope).

Depending on the extent of the kink and on how much the rope rises
during the kink, one can expect the two legs of the tube to hit each
other at somewhere between right angles and an antiparallel orientation.
\inlinecite{Linton&al2001} explored how reconnection would happen at
high twist ($\Phi=20\pi$) for such collisions. Considering an idealized
set-up with periodic boundary conditions and magnetically isolated,
straight twisted flux tubes brought together by a prescribed slow
stagnation-point flow, these authors found that antiparallel flux tubes
of like helicity bounce off each other with hardly any reconnection
occurring, hence they termed  this case ``bounce interaction.'' This is
a natural consequence of the fact that the fields at the surface of the
flux tubes, which are dominated by the  azimuthal component for such
strong twist, are approximately parallel where the tubes come into
contact. They found a similar bounce interaction when the tubes collide
at right angles with the same signs of twist helicity and crossing
helicity (i.e., RR2: right handed tubes at a collision angle of $2\pi/8$
or LL6: left handed tubes at an angle of $6\pi/8$, in the nomenclature
of \opencite{Linton&al2001}). On the other hand, they found that the
tubes undergo significant reconnection at right angles, and in fact
tunnel through each other, if the twist helicity and crossing helicity
are of opposite signs (RR6 or LL2). \inlinecite{Linton2006} then showed
that this tunnel interaction can occur for such collisions at twists as
low as $\Phi=9\pi$, which equals the twist of the kink simulation shown
in Figures~\ref{f:rope_fl1}--\ref{f:rope_iso_j} here. This raises the
interesting question of whether such a tunnel reconnection could occur
in the kink-induced reconnection simulations we present here.
Unfortunately, it turns out this can not happen, as the kink always
gives a crossing number which is unfavorable for tunnel reconnection.
The kink instability in a left hand twisted flux tube induces left hand
writhing, which in turn gives a left handed flux tube crossing. Thus the
collision of flux tube legs induced by the kink instability gives the
same sign of crossing number and twist (LL6 or RR2), and can not induce
tunneling reconnection.

In spite of this, we do find significant reconnection in this
kink-induced collision. This, we interpret, is due to the fact that the
twist here is initially less than half that of the tubes studied by
\inlinecite{Linton&al2001}, and the kink and eruption of the tubes
subsequently lower it even further. This allows the toroidal field to
dominate over the twist field at the point where the tubes collide. The
toroidal field is antiparallel for an antiparallel tube collision, and
is at right angles for a right angle collision. Both orientations are
favorable for reconnection, and so significant flux tube reconnection is
generated by these collisions. Furthermore, the reconnection is here
driven and enforced by the overall helical deformation of the flux rope,
which is a fast process, since the helical kink is an ideal MHD
instability. The current channels of diameter $2a\approx0.4$ merge
completely within $\approx\!8\tau_A$ (between $t\approx12.5$ and
$t\approx20$), corresponding to a high reconnection inflow Alfv\'en Mach
number of $\approx\!0.05$.

A further difference to the simulations in \inlinecite{Linton&al2001} is
given by the presence of the helical current sheet. This sheet and the
current flowing in it wrap around the current channel in a helical
manner, with the overall direction in the sheet being opposite to the
direction of current flow in the channel. When the two layers of the
helical current sheet are squeezed together between the approaching,
nearly vertical legs of the channel, they are nearly antiparallel (see
the third panel of Figure~\ref{f:rope_iso_j}). Such a double current
sheet is known to be susceptible to the double tearing mode. Previous
two-dimensional simulations of this mode had sufficient spatial
resolution to demonstrate that it
creates a chain of magnetic islands in each layer, with the island
growth in each layer facilitating the growth in the other layer. This
leads to fast reconnection which depends only weakly on the resistivity
(\eg, \opencite{Pritchett&al1980}; \opencite{Matthaeus&Lamkin1986};
\opencite{Wang&al2007}). As in a single current sheet, the islands in
each layer merge subsequently, driven by the coalescence instability.
The merged islands---plasmoids---typically oscillate and are eventually
ejected at high speed along the sheet if no symmetry is prescribed
\cite{Tajima&al1987, Schumacher&Kliem1996, Magara&al1997, Barta&al2008}.
Such acceleration of plasmoids along the sheet is amplified in the
double tearing mode by the repelling force between neighboring plasmoids
in different layers because the currents are oppositely directed
\cite{Matthaeus&Lamkin1986}. Hence, the double tearing mode has the
potential to support fast reconnection, involving plasmoid formation and
ejection, when two layers of the helical current sheet are squeezed
between the approaching legs of a kinking flux rope.

Far higher spatial resolution than is available in our simulation would be
required to resolve the details of the reconnection between the two layers of
the helical current sheet and between the flux rope legs in three dimensions.
Therefore, we cannot here model the double tearing mode or other processes that
may form plasmoids or other structures capable to confine the accelerated
particles in compact radiation sources, as observed in the 18 April 2001 flare.
Rather, we have to rely on the basic knowledge reviewed above. It must be kept
in mind, however, that the rapid driving of the reconnection in the double
current sheet by the approaching flux rope legs may modify the evolution of the
double tearing mode with respect to the quoted numerical simulations, which
considered a double current sheet without external driving.

The reconnection induces a peak parallel electric field drop equal to
the rate of flux change \cite{Hesse&al2005}, which is directly available
to particle acceleration, and a perpendicular electric field of order
$E_\perp=0.05V_AB$, part of which can be tapped by particles drifting in
the inhomogeneous reconnection volume (\eg,
\opencite{Ambrosiano&al1988}; \opencite{Kliem1994};
\opencite{Drake&al2006}; \opencite{Karlicky&Barta2007}). Assuming, for
example, a field strength $B=100$~Gauss and a particle density
$n=10^{10}$~cm$^{-3}$, we have $V_{A0}\approx2000$~km\,s$^{-1}$ and
$E_\perp\sim1$~kV\,m$^{-1}$. Assuming in addition an initial flux rope
height of 30~Mm, the simulation results scale to a flux in the current
channel of $\sim\!10^{19}$~Mx (a factor $\lesssim\!10$ lower than the
flux in the whole flux rope), reconnected in $\sim\!110$~s, which yields
a parallel potential of $\sim\!1$~GV. Both are huge values, more than
sufficient to enable the acceleration of particles to the observed
energies.

The region of reconnection between the flux rope legs does not participate in
the fast rise of the flux rope's upper part, rather it remains in the range of
heights where it commences, $z\lesssim1$, with a slight tendency to shift
downwards from the legs' initial touching point
(Figures~\ref{f:shape(pi)}--\ref{f:rope_iso_j}). This is in qualitative
agreement with the observations of the 18 April 2001 eruption described in
Paper~I, which indicate that leg-leg reconnection commenced at about the time
the crossing point of the legs had appeared at the limb, \ie, at a height
slightly above $\sim0.1R_\odot$, and that the region of interaction
subsequently remained near the limb or even retracted slightly (see also
Figure~\ref{f:comparison1} below). However, the heights of the observed
crossing point and of the commencing reconnection in the simulation differ by a
factor $\sim\!1.5$. This value can roughly be estimated from the length of the
polarity inversion line between the main flux concentrations in AR~9415 before
it rotated beyond the limb, about $10^\circ$ or $\sim\!0.2R_\odot$, which
should be comparable to the distance, $D$, between the footpoints of the
erupting flux loop. The observationally indicated height range of leg-leg
reconnection thus is $h_\mathrm{r}\lesssim D/2$, while the simulation above,
with $D=3.3$, yields $h_\mathrm{r}\lesssim D/3$. This moderate disagreement can
be resolved by the inclusion of a second process that causes the flux to rise
higher before it starts to reconnect. Candidate processes are an accelerated
slow-rise phase, for example due to slow photospheric motions which lead to an
amplifying inflation of the current-carrying coronal field (\eg,
\opencite{Torok&Kliem2003}; \opencite{Aulanier&al2010}), as well as any other
CME driver, for example another instability. We consider the latter possibility
in the following subsection.

High twist and leg-leg reconnection do not always lead to ejective
behavior. When the simulation is repeated with a stronger line current
(as in \opencite{Torok&Kliem2005}) but otherwise identical parameters, a
failed eruption is found. Reconnection commences between the approaching
legs of the rope, as decribed above, but soon thereafter it also
commences in the section of the helical current sheet above the flux
rope apex, where the current density is now much higher due to stronger
flux pileup. The latter reconnection, described in
\inlinecite{Torok&Kliem2005}, cuts the rope at its top, thus preventing
an ejective evolution.

\subsection{Helical Kink Supported by the Torus Instability}
\label{ss:hKI+TI}

\begin{figure}[t]                                                
\centering
\includegraphics[width=\textwidth]{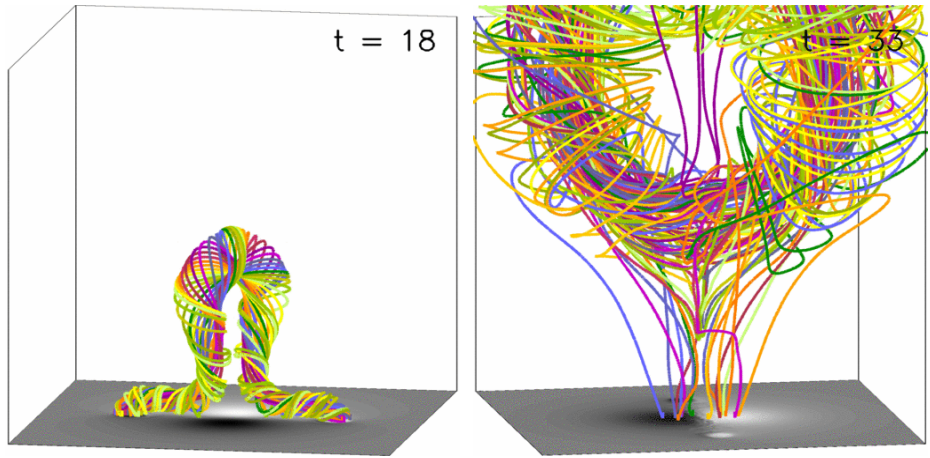}  
\caption{
{\it Left:} Field lines of a kink and torus unstable flux rope of initial twist
$\Phi=8\pi$ just before leg-leg reconnection commences, showing that the height
range of leg approach is somewhat enlarged in comparison with the case shown in
Figures~\ref{f:rope_fl1}--\ref{f:rope_iso_j}, due to the additional upward
acceleration by the torus instability. {\it Right:} The same configuration at a
later time showing that part of the flux in the escaping upper part of the
original flux rope produces closed loops as a result of the reconnection
between the legs, while the other part has reconnected with the ambient field.
The field line start points lie in a circle of radius $\approx\!(2/3)a(t)$
centered at the flux rope's apex. The inner subvolume of the box of size $6^3$
is shown. The views differ by a rotation of $70^\circ$ about the $z$ axis.}
\label{f:hKI+TI}
\end{figure}

\begin{figure}[t]                                                
\centering
\includegraphics[width=.7\textwidth]{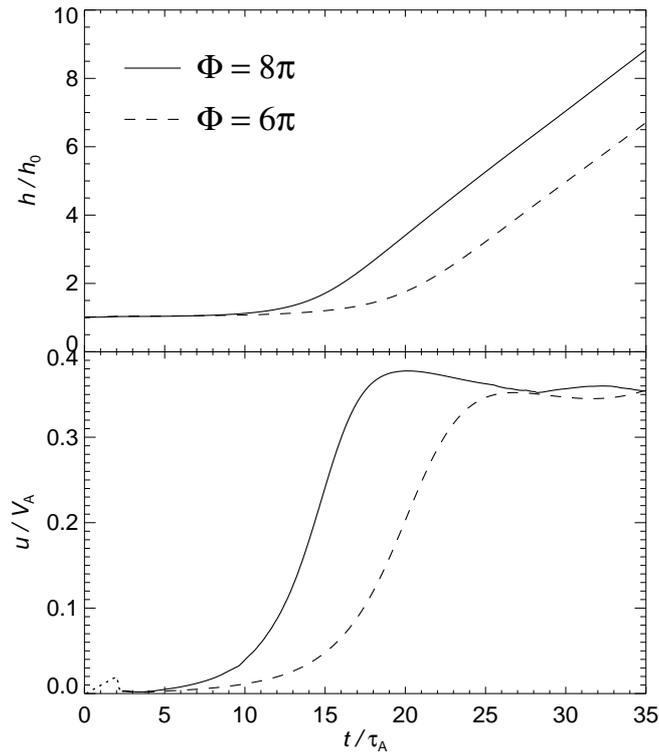}
\caption{
Rise profiles of the fluid element at the flux rope apex for the two
simulations of Section~\ref{ss:hKI+TI} that are compared to the 17~GHz images
of the flare on 18 April 2001 in Figure~\ref{f:comparison1} below \emph{(solid
lines)} and in Figure~9 of Paper~I \emph{(dashed lines)}. The initial velocity
perturbation applied at the apex is shown dotted.}
\label{f:rise}
\end{figure}

\begin{figure}[t]                                                
\centering
\includegraphics[width=.6\textwidth]{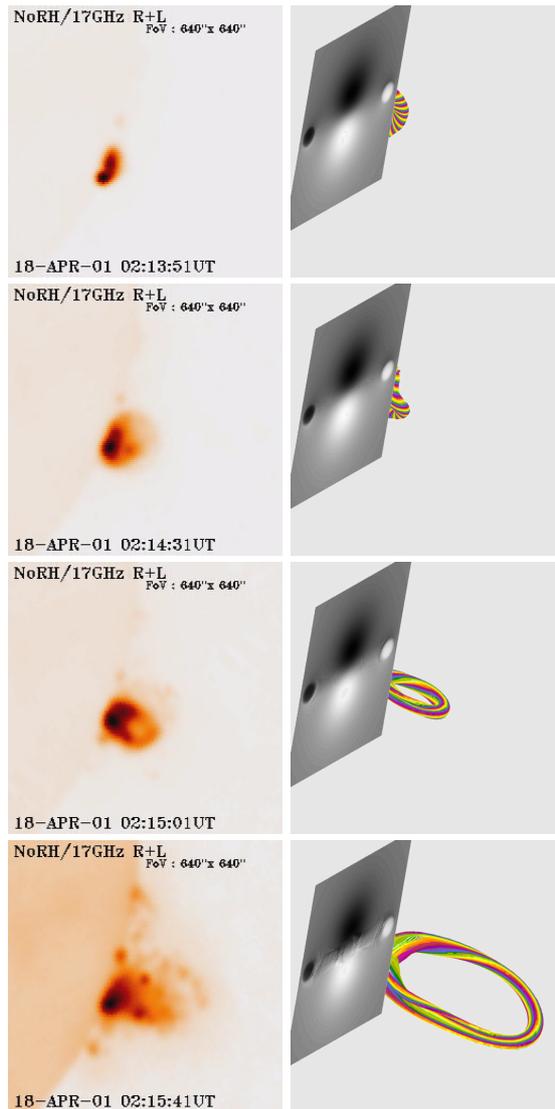} 
\caption{
Comparison of the evolving 17~GHz source in the 18 April 2001 eruption with the
shape of the kink-unstable flux rope of initial twist $\Phi=8\pi$ shown also in
Figure~\ref{f:hKI+TI}. The simulated flux rope is displayed at the same viewing
angle as the partly occulted microwave source (tilted away from the observer by
26$^\circ$, so that the magnetogram, $B_z(x,y,0,t)$, is seen from below the
simulation box). The field lines outline the core of the flux rope at $t=0$,
12, 17, and $24\tau_A$.}
\label{f:comparison1}
\end{figure}

Here we describe a flux rope that is susceptible to the helical kink
mode and to the torus instability from the beginning. The torus
instability \cite{Kliem&Torok2006} is the process behind the well-known
catastrophe of a flux rope-arcade configuration
\cite{vanTend&Kuperus1978, Forbes&Isenberg1991, Lin&al1998}. It provides
an additional upward acceleration of the flux rope, raising the height
range of leg-leg reconnection due to the helical kink
mode. In order to make the initial equilibrium considered in
Section~\ref{ss:R_hKI} also torus unstable, one has to modify the
external poloidal field such that it falls off more rapidly with height,
or to reduce the stabilizing influence of line-tying. Let us first
describe a simulation that realizes the former option by reducing the
distance of the external poloidal field sources to $L=0.25$. The twist
is reduced slightly to bring the time scales of the two instabilities
closer together (so that the torus instability can influence the flux
rope sufficiently early in the process of helical deformation), but it
is still kept at a rather high value of $\Phi=8\pi$. The resulting
evolution of the flux rope is very similar to the case described in
Section~\ref{ss:R_hKI}, except that the upper part of the rising loop is
now stretched upward somewhat stronger, as shown in the left panel of
Figure~\ref{f:hKI+TI}. The reconnection commences at somewhat greater
heights in this run, in the range $z\approx1$, and the current sheets
between the flux rope legs are squeezed in the height range up to
$z\approx1.5$. This remains the range of reconnection between the legs
as the eruption evolves, as indicated by the shape of the field lines in
the right panel of Figure~\ref{f:hKI+TI}. This figure also proves the
leg-leg reconnection by showing that part of the flux in the upper part
of the broken rope closes with itself. The other part reconnects with
the ambient field, forming new, less sharply defined legs.

As in the run in Section~\ref{ss:R_hKI}, the rise profile of the flux rope
apex shows the two phases typical of fast CMEs: a rapid exponential rise to the
peak velocity of nearly $0.4V_A$ is followed by a linear further rise with very
slowly decreasing velocity (Figure~\ref{f:rise}). The main part of the
original flux in the legs of the rope
reconnects in the interval $t\approx\!(18\mbox{--}25)\tau_A$, \ie, shortly
\emph{after} the peak acceleration of the ejection, similar to the run in
Section~\ref{ss:R_hKI} and exactly as inferred for the 18 April 2001 flare (see
Paper~I). Reconnection involving the newly formed legs and the ambient
field continues subsequently at a lower rate until the run is terminated.

This run reproduces the key features of the evolving shape of the microwave
source in the 18 April 2001 flare. Figure~\ref{f:comparison1} displays, from
top to bottom, the top section of a flux loop, the transition to concave-upward
shape, the initially slim ellipse, and finally the considerably expanded and
presumably further rotated ellipse, whose bottom stays close to the limb.

An even better agreement with the observed shape was obtained in another
run which started from a less twisted rope ($\Phi=6\pi$), thus reducing
the growth rate of the helical kink further and permitting a stronger
influence of the torus instability. The external poloidal field in this
run was chosen as in Section~\ref{ss:R_hKI} and the torus instability
can grow because the line-tying is reduced by placing the rope at a
higher position ($R=1.1$, $d=0.1$), giving it a nearly semicircular
initial shape. The comparison with the microwave images is shown in
Figure~9 of Paper~I in the same format as in Figure~\ref{f:comparison1}
here. The rise profile is included in Figure~\ref{f:rise}. It shows the
same two-phase velocity profile and the same timing of reconnection
between the flux rope legs ($t\gtrsim23\tau_A$) relative
to the main acceleration phase as the previous two runs.

This third run in particular brings observation and
simulation in reasonable quantitative agreement, resolving the
moderate discrepancy in the height of the crossing/reconnection-onset
point discussed in Section~\ref{ss:R_hKI}. As mentioned there, further
possibilities exist to raise the height of leg-leg reconnection onset
in the numerical modeling; however, we refrain from further
pursuing the parametric study of this minor aspect.

\section{Discussion}
\label{s:discussion}

The analysis of the microwave data of the eruptive flare on 18 April 2001 and
the comparison with the above MHD simulations suggest that the
eruption involved a kink-unstable flux rope that had a high twist of
$\Phi\gtrsim6\pi$. This yields a coherent framework to understand the inverse
gamma shape of the rising main microwave source, as well as the triggering and
ejection of superimposed compact sources that likely emerged from the crossing
point and moved upward to the top of the main source.

The inferred twist lies several times above the often quoted threshold
of $2.49\pi$ for the helical kink instability of a line-tied flux rope
\cite{Hood&Priest1981, Einaudi&VanHoven1983}, so that the question
arises how such a high value can occur. It can likely be acquired by the
flux rope in the eruption process itself, by adding current-carrying
flux to the rising and growing rope through reconnection under it. This
addition of flux is well established theoretically (\eg,
\opencite{Lin&Forbes2000}; \opencite{Lin&al2004}) and observationally
(\eg, \opencite{Qiu&al2004}). Resulting twist values far exceeding the
above threshold have been inferred from some interplanetary CMEs
\cite{Gulisano&al2005, Qiu&al2007}, but the corresponding coronal values
remain elusive.

It is also possible that the initial equilibrium had acquired the required
twist. The threshold found by \inlinecite{Hood&Priest1981} and
\inlinecite{Einaudi&VanHoven1983} is the \emph{lowest possible} threshold,
valid only for a uniformly twisted (Gold-Hoyle) flux rope of infinite radial
extent. Several factors influence the threshold of the instability, among them
the embedding of the flux rope in strong external current-free field, the
inhomogeneity of the radial twist profile and the radius of the rope. In
particular, it has been shown by \inlinecite{Baty2001} that the threshold can
lie at twists of $10\pi$ and higher if the radius of the rope becomes smaller
than the pitch length of the twisting field lines. With regard to the acquired
twist, the event on 18 April 2001 appears to be atypical, belonging to a
minority of cases with very high twist. Other candidate cases for such and even
somewhat higher twist are described in \inlinecite{Vrsnak&al1991},
\inlinecite{Romano&al2003}, and in \inlinecite{Cho&al2009}.

\inlinecite{Cho&al2009} suggested that a kinking flux rope with reconnecting
legs caused an eruptive X-class limb flare on 18 August 2004. They found
evidence for the break-up of a kinked loop at the projected crossing point of
the legs. Hot EUV-emitting plasma was produced at low heights up to the
breaking point, presumably by reconnection, while the top part of the loop was
further accelerated and evolved into a CME. An outward moving compact coronal
hard X-ray source appeared at the position and time of the break-up inferred
from EUV and H$\alpha$ images; this source was seen at higher energies than the
well known coronal sources formed at or above the top of soft X-ray loops in
some events (\eg, \opencite{Sui&Holman2003}). These findings are analogous to
the dynamics of the microwave and hard X-ray sources and to the spectral extent
of the hard X-ray source in the 18 April 2001 flare discussed in Paper~I, and
they are in line with the simulations presented above.

The quick reformation of a flux rope below the reconnecting legs of the
original rope can explain the quick reformation of filaments after some
eruptions, for example, in the 19 July 2000 event investigated by
\inlinecite{Romano&al2003}.

Further possible cases of reconnection at or near the crossing point of rising
kinked filaments were presented in \inlinecite{Ji&al2003} and
\inlinecite{Alexander&al2006} for an event on 27 May 2002 and in
\inlinecite{Liu&Alexander2009} for events on 12 June 2003 and 10 November 2004.
These  authors describe hard X-ray sources near the projected crossing point of
the legs in addition to footpoint sources, and the latter two papers suggest
leg-leg reconnection as one possible mechanism of the underlying particle
acceleration. No break-up of the kinked filaments at the crossing point
occurred, however, and the shapes are suggestive of a dominantly single helical
($k'\sim1$) kink. The kinking of twisted flux was modeled for the first and
third of these events in \inlinecite{Torok&Kliem2005} and
\inlinecite{Williams&al2005}, respectively, assuming a twist of $5\pi$ in both
cases, which led to a  $k'\sim1$ kink. It is likely that leg-leg interaction
occurred in the event on 27 May 2002, which showed the clearest association of
the X-ray source with the crossing point, as a consequence of the failed nature
of the eruption and the cut up of the filament at its top. We are preparing a
corresponding simulation study to detail this suggestion. For the other two
events, which were ejective, we believe that reconnection in the helical
current sheet has likely played a role (partly similar to our suggestion here),
but that the conjecture of leg-leg reconnection requires further study and
justification.

\section{Summary and Conclusion}
\label{s:summary}

MHD simulations of the helical kink instability of arched, line-tied
flux ropes demonstrate that the legs of the flux rope reconnect with
each other if perturbations with two helical turns, \ie, normalized
axial wavenumbers $k'\sim2$, are dominant. This requires twists of
$\approx6\pi$ or higher.

Leg-leg reconnection can cause a complete break-up of the original rope.
The upper part of the original rope can evolve into a CME or stay
confined in the corona. Its fate depends on the overlying field, as for
less twisted erupting flux ropes (see \opencite{Roussev&al2003};
\opencite{Torok&Kliem2005}; \opencite{Kliem&Torok2006}). The lower parts
can form a new flux rope connecting the footpoints of the original rope.
The new rope is still considerably twisted. The break-up reconnection is
found to involve reconnection with the ambient flux in all simulations,
resulting in new connections of the rising upper part of the kinked rope
to the footpoints of originally ambient flux. Furthermore, the downward
reconnection outflow pushes the reformed flux rope towards the bottom
plane, where it reconnects and breaks up a second time. The resulting
two major flux bundles, which connect the footpoints of the original
flux rope with the corresponding major ambient flux concentration
(sunspot) of opposite polarity, represent a state close to the potential
field of the given flux distribution in the bottom plane (illustrated,
e.g., in Figs.~3 and 8 of \opencite{Valori&al2010}.) The
simulations also show that the region of reconnection between the legs
of the highly kinked flux rope stays low in the coronal volume and that
this reconnection commences just after the main acceleration of the flux
rope's upper part.

Such dynamics are in overall agreement with the flare on 18 April 2001 studied
in Paper~I, supporting the interpretation that the microwave source was formed
by a flux rope which was destabilized by the helical kink mode at the onset of
the eruption's main acceleration phase and experienced leg-leg reconnection in
the nonlinear development of the instability. The occurrence of leg-leg
reconnection argues against a significant role for the shear field component in
the writhing of the erupting flux because this effect pushes the legs of the
rope apart.

A quantitative difference relative to the observations is
given by the fact that the crossing point of the inverse gamma shaped
microwave source rose to a height of about half the footpoint distance
before the reconnection commenced, while the simulations yield a height
of about one third the footpoint distance if the eruption is solely
driven by the helical kink mode. This indicates that the initial rise,
at least up to the height where the reconnection commenced, involved a
second process in addition to the helical kink instability. The data do
not provide clues to the nature of that process, which could have been
an accelerated slow-rise phase or any of the triggers of fast rise
discussed in current CME models. Simulations that include the torus
instability in addition to the double helical kink mode, as a
representative model, resolve the discrepancy.

The approach of the flux rope legs squeezes sections of the helical
current sheet associated with the kink mode into a temporary double
current sheet between them. This configuration is susceptible to the
double tearing mode, which is known to facilitate fast reconnection and
the formation of moving plasmoids through subsequent island coalescence.
However, the limited resolution of our simulations does not permit us to
determine whether this mode did actually occur. Driven by the kink
instability, the double current sheet and the flux rope legs reconnect
essentially completely and at a high rate, with the Alfv\'en Mach number
of the reconnection inflow reaching $\approx0.05$.

The path of the moving compact microwave and hard X-ray sources along
the upper part of the inverse gamma source and the coalescence of
multiple blobs into a final compact microwave source at the top of
the inverse gamma source in the 18 April 2001 flare are consistent with
plasmoid formation and propagation in the helical current sheet formed
by the $m=1$ kink instability. The alternative assumption of
formation by reconnection in the vertical current sheet of the standard
flare model appears to be incompatible with these observations (see
Paper~I). 

These findings lead us to conclude that the eruptive flare on 18 April
2001 was likely caused---not exclusively but to a large part---by the
helical kink instability of a highly twisted flux rope.

\acknowledgements
We thank M.~Shimojo for the Nobeyama Radioheliograph maps included
in Figure~\ref{f:comparison1} and T.~G.~Forbes for discussions of the
instabilities considered here.
This study was supported by the DFG, by an STFC Rolling Grant, by
NASA Grants NNH06AD58I and NNX08AG44G, and by
Grant 300030701 of the Grant Agency of the Czech Academy of Sciences.
Financial support by the European Comission through the SOLAIRE network
(MTRM-CT-2006-035484) is gratefully acknowledged.
The research leading to these results has received funding from the
European Commission's Seventh Framework Programme (FP7/2007-2013) under
the grant agreement n° 218816 (SOTERIA project, www.soteria-space.eu).

\end{article}
\end{document}